\documentclass[aps, twocolumn, prl,times, superscriptaddress, floatfix]{revtex4-1}
\usepackage{amsmath}
\usepackage{amssymb}
\usepackage{graphicx}
\usepackage{array}
\usepackage{multirow}
\usepackage{mathptmx}

\begin{document}

\title{Absolute intensity normalisation of powder neutron scattering data}
\author{Joseph A. M. Paddison}
\email{paddisonja@ornl.gov}
\affiliation{Neutron Scattering Division, Oak Ridge National Laboratory, Oak Ridge,
Tennessee 37831, USA}

\begin{abstract}
An important property of neutron scattering data is that they can
be normalised in absolute intensity units. In practice, however, such
normalisation is often not performed, since it can be time-consuming
and subject to systematic uncertainties. Here, a straightforward approach
is presented for absolute intensity normalisation of neutron scattering
data from polycrystalline samples. This approach uses the intensity
scale factor obtained from a Rietveld refinement to normalise the
data to the nuclear Bragg profile of the sample. Factors to convert
the Rietveld scale factor into an absolute normalisation factor are
tabulated for constant-wavelength and time-of-flight data refined
using the popular programs Fullprof and GSAS-II. An example of the
application of this method to experimental data is presented. Advantages,
disadvantages, and extensions of this approach to spectroscopic data
are discussed.
\end{abstract}

\maketitle 

\section{Introduction}

Among crystallographic measurements, neutron scattering has the useful
and unique property that the measured scattering profile can be placed
on an absolute intensity scale. This property is important in the
study of disordered crystalline materials using total scattering or
pair-distribution function (PDF) techniques \cite{Toby_1992,Billinge_2004},
where it enables accurate measurement of coordination numbers, and
can be necessary to use modelling approaches such as reverse Monte
Carlo refinement \cite{Keen_1990,Tucker_2007}. It is also important
in magnetic neutron scattering, since it allows for quantitative determination
the magnitude of ordered magnetic moments, diffuse scattering, and
excitations \cite{Ziebeck_1980,Steinsvoll_1984}. This information
cannot typically be obtained from other techniques such as resonant X-ray
scattering, and is relevant to scientific problems of ongoing importance,
such as the question of the ``missing'' magnetic spectral weight
in high-$T_{c}$ superconductors \cite{Lorenzana_2005} and the ``quantumness''
of the magnetic excitations of quantum magnets \cite{Mourigal_2013}. 

Despite the value of absolute intensity measurements, most published
neutron scattering patterns only report the intensity in relative
units (e.g., counts per unit time) rather than absolute units (e.g.,
$\mathrm{barn\thinspace sr^{-1}\thinspace atom^{-1}}$). A likely
explanation is that current approaches to data normalisation present
some practical difficulties. For example, a common approach is to
normalise data to the incoherent scattering from a separately-measured
vanadium calibration sample \cite{Wignall_1987,Xu_2013}. This method
requires accurate measurement of the vanadium standard, the packing
fraction of the sample, the mass of vanadium in the beam, and the
absorption of the sample and the vanadium, which can be time-consuming
\cite{Xu_2013,Soper_2011}. An alternative approach is to normalise
the data to the incoherent scattering from the sample itself \cite{Xu_2013}.
This difficulty here is that extraneous effects -- background scattering
from the apparatus, diffuse scattering from chemical short-range ordering
in the sample, and additional incoherent scattering from any hydrogen-containing
impurities -- will add to the apparent incoherent scattering, giving
a systematic uncertainty that is difficult to quantify \cite{Xu_2013}.

In this short communication, I present an alternative approach to
absolute normalisation of neutron scattering data for polycrystalline
materials: normalisation to the nuclear Bragg profile of the sample
itself. The nuclear Bragg profile is routinely calculated in widely-used
software for Rietveld refinement \cite{Rietveld_1969}, such as Fullprof
\cite{Rodriguez-Carvajal_1993a,Rodriguez-Carvajal_2001}, GSAS-II
\cite{Toby_2013}, and Topas \cite{Coelho_2018}. This calculation
can be done in absolute intensity units for any material with a known
crystal structure, since the relevant scattering equations and scattering
lengths are known \cite{NN_1992}. Unfortunately, however, Rietveld
programs do \emph{not} calculate the nuclear Bragg profile in absolute
intensity units, so the intensity scale factor obtained from a Rietveld
refinement cannot be applied directly to normalise the data. The main
result of this paper is a set of simple formulae (see Table~\ref{tab:results})
that convert the Rietveld scale factor into an ``absolute normalisation
factor'' that transforms the data into absolute intensity units.
Using this approach, scattering data from polycrystalline materials
can be normalised with little effort beyond a Rietveld refinement.

This paper is structured as follows. I start with a derivation of
the relationship between the absolute normalisation factor and the
Rietveld scale factor. I then provide an example using published experimental
data for the frustrated magnet Dy$_{3}$Mg$_{2}$Sb$_{3}$O$_{14}$
\cite{Paddison_2016}. To demonstrate the accuracy and reproducibility
of the method, results are compared for constant-wavelength and time-of-flight
data measured on the same material. I conclude by summarising the
key results and providing a table of absolute normalisation factors
for the common Rietveld programs Fullprof and GSAS-II. A reader who
is interested in applying the approach can skip the derivations and
refer to Table~\ref{tab:results}.

\section{Derivation}

I start by assuming that the nuclear profile intensity measured experimentally,
$I_{\mathrm{expt}}$, is proportional to the absolute nuclear intensity
in units of $\mathrm{barn\thinspace sr^{-1}\thinspace atom^{-1}}$:
\begin{equation}
I_{\mathrm{expt}}=sI_{\mathrm{abs}}.
\end{equation}
Our aim is to find the proportionality constant $s$. We call the
overall intensity scale factor obtained from Rietveld refinement $\mathtt{RietveldScale}$.
This parameter multiplies the intensity $I_{0}$ calculated internally
by the Rietveld refinement program, so that after the refinement has
converged, the result optimally matches the experimental data:
\begin{equation}
I_{\mathrm{expt}}=\mathtt{RietveldScale}\times I_{0}.
\end{equation}
It follows that
\begin{equation}
s=\mathtt{RietveldScale}\times\frac{I_{0}}{I_{\mathrm{abs}}}.\label{eq:abs_scale}
\end{equation}
Hence, to determine $s$, we require expressions for both $I_{0}$
and $I_{\mathrm{abs}}$. These are given below.

The general equation for the $Q$-dependence of the Bragg scattering
intensity for a polycrystalline sample is given by \cite{Squires_1978}
\begin{equation}
I_{\mathrm{abs}}(Q)=\frac{2\pi^{2}}{NV}\sum_{\mathbf{G}}\frac{m_{\mathbf{G}}|F_{\mathbf{G}}|^{2}}{G^{2}}R_{Q}(Q-G),\label{eq:powder_profile_q}
\end{equation}
where $V$ is the volume of the crystallographic unit cell containing
$N$ atoms, $\mathbf{G}$ labels a set of equivalent Bragg reflections
having multiplicity $m_{\mathbf{G}}$, $G=|\mathbf{G}|$ is the length
of a reciprocal-lattice vector, $R_{Q}(Q-G)$ is a resolution function
that obeys the normalisation condition $\int R_{Q}(Q-G)\thinspace\mathrm{d}Q=1$,
and the structure factor
\begin{equation}
F_{\mathbf{G}}=\sum_{j=1}^{N}b_{j}T_{j}\exp\left(\mathrm{i}\mathbf{G}\cdot\mathbf{r}_{j}\right),
\end{equation}
where $\mathbf{r}{}_{j}$ is the position of atom $j$ in the unit
cell, $b_{j}$ is its nuclear scattering length (in units of $10^{-12}$\,cm),
$T_{j}$ is its Debye-Waller factor, and the sum is taken over all
$N$ atoms in the unit cell \cite{Squires_1978}. Eq.~(\ref{eq:powder_profile_q})
gives the intensity in absolute units of $\mathrm{barn\thinspace sr^{-1}\thinspace atom^{-1}}$.

The profile intensity calculated internally in Rietveld refinement
depends on the Rietveld program. In the derivation below, I will use
FullProf \cite{Rodriguez-Carvajal_2001} as an example; the results
for GSAS-II are different and given in Table~\ref{tab:results}.
The profile intensity calculated internally by FullProf is given by
\begin{equation}
I_{0}(x)=\sum_{\mathbf{G}}m_{\mathbf{G}}\left|F_{\mathbf{G}}\right|^{2}L_{x}R_{x}(x-x_{\mathbf{G}}),\label{eq:fullprof_intensity}
\end{equation}
where $x$ denotes either scattering angle $2\theta$ for constant-wavelength
diffraction or time $t$ for time-of-flight diffraction, and $R_{x}(x-x_{\mathbf{G}})$
is a resolution function that again obeys $\int R_{x}(x-x_{\mathbf{G}})\thinspace\mathrm{d}x=1$.
The expressions for Lorentz factor $L_{x}$ and resolution function
are written with subscript $x$ to emphasise that they differ for
time-of-flight \emph{vs.} constant-wavelength data. I note that Eq.~(\ref{eq:fullprof_intensity})
assumes that the data have already been corrected for absorption (and
any other relevant corrections to the overall intensity) before being
used as input for Rietveld refinement.

\begin{table*}
\begin{centering}
\begin{tabular}{c|c|c}
\hline 
 & Fullprof & GSAS-II\tabularnewline
\hline 
\hline 
\multirow{2}{*}{Constant wavelength} & \multirow{2}{*}{$\mathtt{FullprofScale}\times\frac{2\pi^{2}NV}{45\lambda^{3}}$} & \multirow{2}{*}{$\mathtt{GSASScale}\times\frac{2\pi^{2}N}{4500\lambda^{3}}$}\tabularnewline
 &  & \tabularnewline
\multirow{2}{*}{Time of flight} & \multirow{2}{*}{$\mathtt{FullprofScale}\times\frac{4\pi NV\sin\theta}{\mathtt{dtt1}}$
($^{\ast}$)} & \multirow{2}{*}{$\mathtt{GSASScale}\times\frac{4\pi N\sin\theta}{\mathtt{DIFC}}$}\tabularnewline
 &  & \tabularnewline
\hline 
\end{tabular}
\par\end{centering}
\caption{\label{tab:results}Factors to convert measured intensity profile
to absolute intensity units.\emph{ }\textbf{To convert data into absolute
intensity units of $\mathrm{barn\thinspace sr^{-1}\thinspace atom^{-1}}$,
divide the measured intensity profile by the relevant number obtained
from this table. }$\mathtt{FullprofScale}$ is the intensity scale
factor refined using FullProf, $\mathtt{GSASScale}$ is the intensity
scale factor refined using GSAS-II, $V$ is the volume of the unit
cell in \AA$^{3}$, and $N$ is the number of atoms in the
unit cell. For constant-wavelength data, $\lambda$ is the neutron
wavelength in \AA. For time-of-flight data, $\theta$ is
the constant scattering angle of the detector bank, and $\mathtt{dtt1}$
or $\mathtt{DIFC}$ is the diffractometer constant with units of $\mu\mathrm{s}\thinspace\textrm{\AA}^{-1}$
that relates $d$-spacing (in \AA) to time-of-flight (in
$\mu\mathrm{s}$). Note that $N$ here denotes the total number of
atoms per unit cell, so absolute intensity will be in $\mathrm{barn\thinspace sr^{-1}\thinspace atom^{-1}}$.
To normalise to the number of magnetic atoms, replace $N$ by the
number of magnetic atoms per unit cell.\protect \\
\textbf{Important note:} The correction factors in this table assume
that the Rietveld program does not apply overall intensity corrections
for experimental effects such as absorption. If such corrections are
needed, they should be applied directly to the data prior to the Rietveld
refinement.\protect \\
($^{\ast}$) Note: For some formats of time-of-flight data, such as
multi-bank data in RAL format, FullProf internally multiplies the
data by a factor of 1000. In such cases, this conversion factor should
be replaced by $\mathtt{FullprofScale}\times0.001\times\frac{4\pi NV\sin\theta}{\mathtt{dtt1}}$
when applied to the original data.}
\end{table*}

\subsection*{Constant-wavelength diffraction}

To compare the equations for $I_{0}$ and $I_{\mathrm{abs}}$, it
is necessary to change the variable from $Q$ to scattering angle.
For constant-wavelength diffraction, this conversion is
\begin{equation}
Q=\frac{4\pi\sin\theta}{\lambda},\label{eq:bragg_law}
\end{equation}
where $\lambda$ is the fixed incident neutron wavelength in Å, and
$2\theta$ is scattering angle in degrees. The change-of-variable
rule gives
\begin{align}
R_{Q}(Q-G) & =\left|\frac{\mathrm{\mathrm{d}(2\theta)}}{\mathrm{d}Q}\right|R_{2\theta}(2\theta-2\theta_{\mathbf{G}})\\
 & =\frac{90\lambda}{\pi^{2}\cos\theta_{\mathbf{G}}}R_{2\theta}(2\theta-2\theta_{\mathbf{G}}).\label{eq:cw_cov}
\end{align}
It follows from Eqs.~(\ref{eq:powder_profile_q}) and (\ref{eq:cw_cov})
that the absolute scattering intensity (in $\mathrm{barn}\thinspace\mathrm{sr^{-1}\thinspace atom^{-1}}$)
is given by
\begin{equation}
I_{\mathrm{abs}}(2\theta)=\frac{45\lambda^{3}}{2\pi^{2}NV}\sum_{\mathbf{G}}m_{\mathbf{G}}\left|F_{\mathbf{G}}\right|^{2}L_{2\theta}R_{2\theta}(2\theta-2\theta_{\mathbf{G}}),\label{eq:abs_2th}
\end{equation}
where the geometrical factors have been combined into the Lorentz
factor for constant-wavelength diffraction,
\begin{equation}
L_{2\theta}\equiv\frac{1}{2\sin^{2}\theta_{\mathbf{G}}\cos\theta_{\mathbf{G}}}.
\end{equation}
Hence, from Eqs.~(\ref{eq:abs_scale}), (\ref{eq:fullprof_intensity})
and (\ref{eq:abs_2th}) , we obtain the absolute normalisation factor
for FullProf,
\begin{equation}
s_{2\theta}^{\mathrm{Fullprof}}=\mathtt{FullprofScale}\times\frac{2\pi^{2}NV}{45\lambda^{3}}.\label{eq:abs_sf_2th_fp}
\end{equation}
The result for GSAS-II differs because its intensity calculation divides
Eq.~(\ref{eq:fullprof_intensity}) by $V$, and uses units of centidegrees
for scattering angle \cite{Larson_1985}. Following the steps above,
we obtain
\begin{equation}
s_{2\theta}^{\mathrm{GSAS}}=\mathtt{GSASScale}\times\frac{2\pi^{2}N}{4500\lambda^{3}}.\label{eq:abs_sf_2th_gsas}
\end{equation}
To obtain the absolute scattering intensity, simply divide the experimental
profile by the relevant $s_{2\theta}$ for FullProf or GSAS-II.

\subsection*{Time-of-flight diffraction}

For time-of-flight diffraction, the relationship between $Q$ and
$t$ is
\begin{equation}
Q=\frac{4\pi m_{\mathrm{n}}l\sin\theta}{ht},\label{eq:Bragg_law_TOF}
\end{equation}
which is obtained from Eq.~(\ref{eq:bragg_law}) by applying the
de Broglie relation, $\lambda=ht/m_{\mathrm{n}}l$, where $t$ is
time of flight, $l$ is total neutron path length, and $m_{\mathrm{n}}$
is neutron mass. In time-of-flight diffraction, $l$ and the scattering
angle $2\theta$ are fixed for a particular detector bank. Applying
the change-of-variable rule,
\begin{align}
R_{Q}(Q-G) & =\left|\frac{\mathrm{\mathrm{d}}t}{\mathrm{d}Q}\right|R_{t}(t-t{}_{\mathbf{G}})\\
 & =\frac{4\pi m_{\mathrm{n}}l\sin\theta}{hG^{2}}R_{t}(t-t{}_{\mathbf{G}}).\label{eq:tof_cov}
\end{align}
It follows from Eqs.~(\ref{eq:powder_profile_q}) and (\ref{eq:tof_cov})
that the absolute intensity is given by
\begin{eqnarray}
I_{\mathrm{abs}}(t) & = & \frac{m_{\mathrm{n}}l}{2\pi hNV}\sum_{\mathbf{G}}m_{\mathbf{G}}\left|F_{\mathbf{G}}\right|^{2}L_{t}R_{t}(t-t{}_{\mathbf{G}}),
\end{eqnarray}
where the geometrical factors have been combined into the Lorentz
factor for time-of-flight diffraction \cite{Zhang_2023},
\begin{equation}
L_{t}\equiv d^{4}\sin\theta,\label{eq:Lorentz_TOF-1}
\end{equation}
where $d=2\pi/G$ is the $d$-spacing of the Bragg reflection. This
yields for the absolute normalisation factor
\begin{equation}
s_{t}^{\mathrm{Fullprof}}=\mathtt{FullprofScale}\times\frac{4\pi NV\sin\theta}{\mathtt{DIFC}},\label{eq:fullprof_tof}
\end{equation}
where
\begin{equation}
\mathtt{DIFC}=\frac{2m_{\mathrm{n}}l\sin\theta}{h}
\end{equation}
is an instrument parameter (also called $\mathtt{dtt1}$ in FullProf),
which has units of $\mu\mathrm{s}\thinspace\textrm{\AA}^{-1}$ and relates
$d$-spacing in \AA~to time-of-flight in $\mu\mathrm{s}$.
For GSAS, the result is
\begin{equation}
s_{t}^{\mathrm{GSAS}}=\mathtt{GSASScale}\times\frac{4\pi N\sin\theta}{\mathtt{DIFC}}.
\end{equation}
Again, to obtain the absolute scattering intensity, simply divide
the experimental profile by the relevant $s_{t}$ for FullProf or
GSAS-II.

\section{Worked Example}

\begin{figure}
\includegraphics[scale=1]{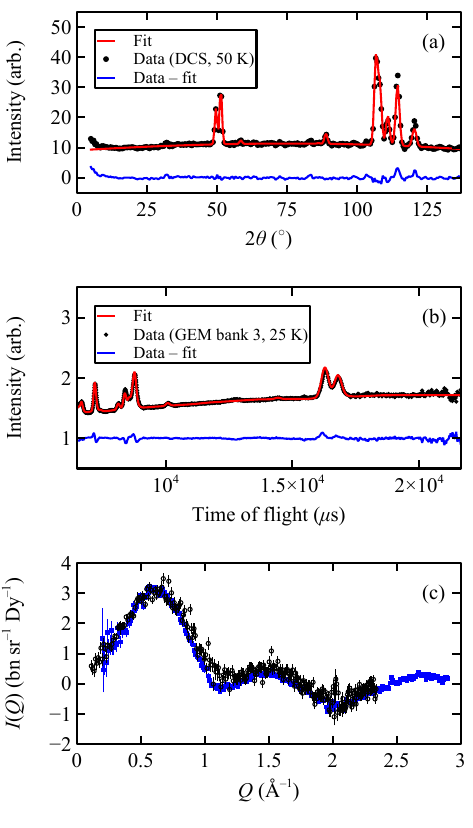}\caption{\label{fig:example_fig}(a) Nuclear refinement of constant-wavelength
neutron data ($\lambda=5\,\text{Å}$) measured on Dy$_{3}$Mg$_{2}$Sb$_{3}$O$_{14}$
at $T=50$\,K using the DCS spectrometer at NIST, showing data (black
circles), Rietveld fit (red line) and difference (blue line). (b)
Nuclear refinement of time-of-flight neutron data measured on Dy$_{3}$Mg$_{2}$Sb$_{3}$O$_{14}$
at $T=25$\,K using the GEM diffractometer at ISIS, showing data
(black circles), Rietveld fit (red line) and difference (blue line).
(c) Magnetic diffuse scattering for Dy$_{3}$Mg$_{2}$Sb$_{3}$O$_{14}$
obtained in absolute intensity units using formulae given in Table~\ref{tab:results},
showing data at $T=0.5$\,K measured on DCS (black circles) and GEM
(blue squares). In each case, a higher-temperature data set (25 or
50\,K) has been subtracted from the 0.5\,K data to remove nonmagnetic
scattering. Data were previously published in~\cite{Paddison_2016}.}
\end{figure}

I now discuss an application of this method using experimental neutron-scattering
data for the frustrated magnetic material Dy$_{3}$Mg$_{2}$Sb$_{3}$O$_{14}$,
which exhibits strong magnetic diffuse scattering. These data were
previously published in~\cite{Paddison_2016}, but the normalisation
was not discussed in detail. Constant-wavelength data were collected
using the DCS spectrometer at NIST \cite{Copley_2003}; the data
were integrated over energy and can be treated as diffraction data.
Time-of-flight data were collected using the GEM diffractometer at
ISIS \cite{Hannon_2005}. 

This material crystallises in the $R\bar{3}m$ space group and the
magnetic Dy atoms occupy the $9e$ site. Rietveld refinements were
performed using FullProf to DCS data measured at 50\,K with $\lambda=5\thinspace\text{Å}$
(see~\cite{Paddison_2016} for details). A reasonable fit is
obtained {[}Fig.~\ref{fig:example_fig}(a){]}. The refined unit-cell
volume was 799.6(2)\,\AA$^{3}$ and the Rietveld scale factor
was 0.071(1), indicating a statistical error $\sim$2\%. Using these
values in Eq.~(\ref{eq:abs_sf_2th_fp}), we obtain $s_{2\theta}^{\mathrm{Fullprof}}=1.80(3)$.
Rietveld refinements were also performed to GEM data measured at 25\,K
{[}Figure~(\ref{fig:example_fig})(b){]}, including data from banks
1 to 4 \cite{Paddison_2016}. Taking bank 1 as an example, $2\theta=9.39^{\circ}$
and $\mathtt{DIFC}=2814.83$\,$\mu\mathrm{s}\thinspace\textrm{\AA}^{-1}$
, the refined unit-cell volume was 799.153(4)\,\AA$^{3}$,
and the Rietveld scale factor was 0.00516(5). Using these values
in Eq.~(\ref{eq:fullprof_tof}), we obtain $s_{t}^{\mathrm{Fullprof}}=0.0513(5)$. 

To isolate the correlated magnetic diffuse scattering, which develops
at temperatures below 25\,K, the 25\,K or 50\,K data were subtracted
from low-temperature data collected at 0.5\,K on each instrument.
The result was divided by the appropriate $s^{\mathrm{Fullprof}}$
determined as above, to place both data sets into absolute units of
$\mathrm{barn\thinspace sr^{-1}\thinspace Dy^{-1}}$. For the GEM
data, the different banks were also merged. The results shown in Fig.~\ref{fig:example_fig}(c)
demonstrate excellent agreement between the normalised diffuse scattering
measured on GEM and DCS. This result suggests that the accuracy of
this approach is likely to be significantly better than the $\sim$$20$\%
systematic uncertainty that has been suggested as typical of other
normalisation methods \cite{Xu_2013}.

\section{Discussion and conclusions}

I have presented a set of factors that allow neutron scattering data
of polycrystalline materials to be placed on an absolute intensity
scale using the Rietveld scale factor. This method offers three main
advantages. First, since the nuclear scattering of the sample is used
as an internal normalisation standard, no additional measurements
are required. Second, systematic errors are likely to be reduced compared
with approaches that use measurements of an external standard. Third,
since Rietveld refinement is already an essential part of data analysis
for many systems, data normalisation can be done with little extra
effort. 

The main limitation of this approach it that it is not applicable
to liquids, amorphous materials, or polycrystalline materials where
the average structure is not well modelled using Rietveld refinement.
As such, it is most useful for magnetic materials with well-ordered
crystal structures, and to materials where a distinct average crystal
structure coexists with short-range order. Notably, the latter class
of systems includes many functional materials studied using PDF approaches
\cite{Toby_1992,Peterson_2021}, where both the Bragg profile and
the PDF are often used to inform atomistic models \cite{Tucker_2007}.

The approach outlined here can also be applied to neutron spectroscopy
data collected on polycrystalline samples, e.g., for measurements
of phonons or magnetic excitations. In this case, data collected on
direct-geometry spectrometers are binned on a grid of $2\theta$ \emph{vs.} energy transfer, integrated over the elastic energy resolution, and the energy-integrated
data are used as input to a Rietveld refinement. The entire spectrum
can then be normalised (in $\mathrm{barn\thinspace sr^{-1}\thinspace atom^{-1}}$
per unit energy) using the factors given in Table~\ref{tab:results};
some examples are discussed in~\cite{Paddison_2024a,Paddison_2024}. I therefore
hope that this work will facilitate wider use of the quantitative
intensity information available from neutron diffraction and spectroscopy data.

\acknowledgments{I am grateful to Stuart Calder, Danielle Yahne, and Ross Stewart for
encouraging me to write this paper. Initial work was supported by
the College of Sciences of Georgia Institute of Technology. Manuscript
preparation was supported by the U.S. Department of Energy, Office
of Science, Scientific User Facilities Division.}


\end{document}